\documentclass{emulateapj}

\newcommand{\hdnum}{HD\,176582} 

\slugcomment{Submitted to the Astronomical Journal 22 December 2010; Accepted 27 February 2011}

\shorttitle{Magnetic Field and Magnetosphere of \hdnum}
\shortauthors{Bohlender \& Monin}

\begin{document}

\title{The discovery of a strong magnetic field and co-rotating magnetosphere in the helium-weak star \hdnum\thanks{Based on observations obtained using the 
Dominion Astrophysical Observatory, Herzberg Institute of 
Astrophysics, National Research Council of Canada and the Canada-France-Hawaii Telescope (operated by the National Research 
Council of Canada, the Centre National de la Recherche 
ScientiÞque of France, and the University of Hawaii).}}

\author{David A. Bohlender and Dmitry Monin}
\affil{National Reseach Council of Canada, Herzberg Institute of Astrophysics, \\
5071 West Saanich Road, Victoria, BC, Canada V9E 2E7}
\email{david.bohlender@nrc-cnrc.gc.ca}

\begin{abstract}
We report the detection of a strong, reversing magnetic field and variable H$\alpha$ emission in the bright helium-weak star \hdnum\/ (HR\,7185).   
Spectrum, magnetic and photometric variability of the star are all consistent with a precisely determined period of $1.^{\!\!\rm{d}}5819840 \pm 0.^{\!\!\rm{d}}0000030$
which we assume to be the rotation period of the star.
From the magnetic field curve, and assuming a simple dipolar field geometry, we derive a polar field strength of approximately 7\,kG and a lower limit  of 52\degr\/ for the inclination of the rotation axis.  
However, based on the behaviour of the H$\alpha$ emission we adopt  a large inclination angle of 85\degr\/ and this leads to a large magnetic obliquity of 77\degr.
The H$\alpha$ emission arises from two distinct regions located at the intersections of the magnetic and rotation equators and  which corotate with the star at a distance of about $3.5\,R_*$ above its surface. 
We estimate that the emitting regions have radial and meridional sizes on the order of $2\,R_*$ and azimuthal extents (perpendicular to the magnetic equator) of less than approximately $0.6\,R_*$.
\hdnum\ therefore appears to show many of the cool magnetospheric phenomena as that
displayed by other magnetic helium-weak and helium-strong stars such as the prototypical helium-strong star $\sigma$\,Ori\,E.
The observations are consistent with current models of magnetically confined winds and rigidly-rotating magnetospheres for magnetic Bp stars.
\end{abstract}


\section{Introduction}

\objectname[HD 176582]{\hdnum} (HR\,7185 = V545 Lyr) is a relatively bright, chemically peculiar (Bp) B5\,IV star that has, somewhat surprisingly, been rather poorly studied given its convenient northern declination.
 \citet{nissen74} first suggested
that the star is a member of the helium-weak class of objects.  Subsequent narrow-band
photometric measurements of the strength of the \ion{He}{1} $\lambda$4026 line \citep{pedersen76}
provided the first evidence for spectrum variability with a possible period of $0.^{\!\!\rm{d}}814$. 
\citet{catanzaro99} confirmed both the spectrum and photometric variability of the star using spectra of the
\ion{He}{1} $\lambda$5876 line and Hipparcos photometry \citep{esa97}.
However they could not confirm the \citet{pedersen76} period but instead found $\rm{P}=1.^{\!\!\rm{d}}58175$.

Our interest in \hdnum\/ arose from the discovery that the star is a radio source at 3.6\,cm with a VLA flux of $0.46 \pm 0.05$\,mJy measured in 1993 \citep[Drake, private communication;][]{drake06}.
A number of other radio sources among the magnetic helium-peculiar stars (comprised of the B5-B9 helium-weak stars with anomalously weak \ion{He}{1} lines for their temperature and the more massive B1-B3 helium-strong stars with anomalously strong \ion{He}{1} lines) have proven to manifest other observational phenomena likely associated with stellar magnetospheres.
This includes variable H$\alpha$ emission or absorption \citep{short94,townsend05b,wade06,leone10}, modulated satellite UV resonance lines \citep[e.g.][]{shore90} and variations in high-level Balmer lines \citep{groote76,smith07}.
The prototype of the helium-strong stars, $\sigma$\,Ori\,E, is the best studied example \citep[e.g.][]{oksala11}.  
On its $1.^{\!\!\rm{d}}19$ rotation period this star shows a roughly dipolar magnetic field varying between -2 and +3\,kG, variable H$\alpha$ emission and shell absorption apparently caused by the changing orientation of two lobes of circumstellar plasma trapped near the intersection of the star's magnetic and rotation equators and rotating rigidly with the surface, as well as enhanced satellite UV resonance line absorption and photometric ``eclipses'' when the magnetic equator (and the bulk of the circumstellar plasma) crosses the line of sight to the observer.

Anticipating that \hdnum\/ might show similar phenomena, we added it to our H$\alpha$ spectroscopy and H$\beta$ spectropolarimetry observing programs on the Dominion Astrophysical Observatory's (DAO) 1.8\,m Plaskett and 1.2-m telescopes.  In this paper we report the discovery of rotationally modulated H$\alpha$ emission variability in \hdnum\/ and also the detection of the star's strong, approximately dipolar magnetic field with the new spectropolarimeter module on the DAO 1.8\,m telescope.

\section{Observations}

The \hdnum\/ observations discussed in this paper include spectra and spectropolarimetry obtained with the DAO 1.8\,m Plaskett telescope, spectra from the DAO 1.2\,m telescope, two high-resolution, high signal-to-noise (S/N) spectra acquired with ESPaDOnS at the Canada-France-Hawaii Telescope (CFHT) and archival Hipparcos photometry.   
Details are given in the following sections.

\subsection{Spectroscopy}

The majority of the DAO spectroscopic observations of \hdnum\/ were acquired between 2005 August 21 (UT) and 2010 July 14.
Intensive observing began after 2005 August 21; a few spectra were obtained considerably earlier in 1999 March and 2000 August but we did not recognize the spectrum variability until the more intensive 2005 August run.

All of the 1.8\,m spectra (115 in total) were acquired between 1999 March 11 and 2009 May 1 with the Cassegrain spectrograph and 21\,inch camera with the 1800\,g/mm grating used in first order.  
Except for a few early 1200 and 1800\,s observations obtained with the VSIS21R image slicer, data were acquired with a long, 1\,arcsecond slit and exposure times of 300 or 600\,s.
The 10\,\AA/mm dispersion and comparable projected slit widths of the image slicer and slit provided a spectral resolution of $R \approx 15\,000$.
The spectral window includes H$\alpha$ and the \ion{He}{1} $\lambda$6678 line.  A few spectra at other wavelength regions have also been obtained but are not discussed here.

Seventy-two DAO 1.2\,m spectra were obtained between 2000 August 14 and 2010 July 14 with most of the observations obtained after 2005 September 8 when the star's spectrum variability had been recognized in 2005 August 1.8\,m spectra. 
Most of these were obtained with the McKellar spectrograph's long 96\,inch camera and 830\,g/mm mosaic grating in first order with a dispersion of 4.8\,\AA/mm, but on a few nights data were also obtained with the short 32\,inch camera and 1200\,g/mm holographic grating, also operating in first order with a dispersion of 10.1\,\AA/mm.  The IS32R image slicer was used for all data acquisition and provided $R=30\,000$ for the long-camera data and nominally $R=24\,000$, but substantially undersampled spectra, for the short-camera data.
Except for the three early 1200\,s observations obtained in 2000, exposure times were 1800\,s.

All of the DAO spectra were processed in a standard fashion with IRAF\footnote{IRAF is distributed by the National Optical Astronomy Observatory, which is operated by the Association of Universities for Research in Astronomy, Inc., under cooperative agreement with the National Science Foundation.}.  Telluric lines were removed by using very high S/N spectra of the bright, rapidly rotating stars $\alpha$ Leo or $\zeta$ Aql as telluric standards.
To improve the S/N we have applied a 3\,pixel boxcar smoothing to all 10\,\AA/mm spectra and 5\,pixel smoothing to the 4.8\,\AA/mm spectra.

In the course of an unrelated observing program we were also able to obtain two spectra of \hdnum\/ with the ESPaDOnS spectrograph at the CFHT on 2006 June 14 and 18 (UT).  
Exposure times were 600\,s and the echelle spectrograph was used in its ``star-only'' mode to provide $R=80\,000$.
These data were processed with the proprietary data reduction package provided for ESPaDOnS users, Libre-ESpRIT \citep{donati97}.
We have also applied a 3\,pixel boxcar smoothing to the processed ESPaDOnS spectra.

Table \ref{spectra} provides the DAO or CFHT odometer number,  observation date, and rotation phase (see below) of the 189 spectra we have obtained for \hdnum.  
The DAO 1.2\,m, 1.8\,m and CFHT spectra (in the region of H$\alpha$ and after smoothing) have typical S/N per pixel values of 300, 500 and 750 respectively.

\subsection{Spectropolarimetry}

The DAO 1.8\,m Plaskett Telescope was also used to acquire 16 spectropolarimetric observations of \hdnum\/ with
the new DAO spectropolarimeter (Monin et al.~2011, in preparation).
The spectropolarimeter uses the same Cassegrain spectrograph configuration described above except a polarimeter module designed and built at the DAO is installed immediately behind a 2\,arcsecond circular entrance to the spectrograph.

This module consists of the three optical elements: a fixed quarter-wave plate, a switchable half-wave plate, and a beam displacer.
In a magnetic field the Zeeman effect splits spectral lines into multiple components with different polarization states.
The line-of-sight component of the field produces two groups of lines with opposite circular polarizations.
The quarter-wave plate converts these two circularly polarized beams into two orthogonal, linearly polarized beams. 
The beam displacer then separates these by about 2\,mm.
The beams travel parallel to each other inside of the spectrograph and land on different rows on the CCD detector to produce the so-called ordinary and extraordinary spectra, separated by about 40\,pixels.
The switchable half-wave plate is a ferroelectric liquid crystal (FLC);
a voltage applied between its electrodes rotates linearly polarized light by 90\degr.
As a result, the two spectra exchange places as the voltage is switched.
This switching helps to minimize most instrumental effects such as the slightly different optical paths of the two beams, the spectrograph response, pixel-to-pixel CCD sensitivity variations, etc.

The $1752\times532$ SITe-2 CCD with 15\,$\mu$m pixels is used as the detector and at a dispersion of 10\,\AA/mm provides a wavelength coverage of
approximately 260\,\AA\/ centered on H$\beta$.
The original CCD hardware and software were modified so that the charge on the CCD can be shuffled back and forth by the 40\,pixel distance between the ordinary and extraordinary spectra in step with the switching of the FLC plate during each exposure. 
As a result of this back and forth charge shuffle, three spectra are formed on the CCD: the left circularly polarized ordinary spectrum (LO), the left circularly
polarized extraordinary spectrum (LE), and the combined right circularly polarized ordinary plus the right circularly polarized extraordinary spectrum (RO+RE).
We typically use 60 shuffles per exposure.
The CCD's parallel transfer efficiency and charge traps currently limit the number of shuffles to a few hundred per exposure.
To make a single magnetic field observation a series of (usually) 12 to 16 short (e.g.\ 5\,min for \hdnum) exposures is obtained in order to keep the switching time shorter than the typical time of the instrumental response changes.
This also allows an estimate of the magnetic field errors to be made from the scatter of measurements from individual exposures.

Every image undergoes bias subtraction, cosmic ray removal, and background removal using standard ESO-MIDAS routines.
The background removal is carried out for each of the three spectra (LO, LE, and RO+RE) independently but we ensure that the three background levels match where they intersect.
LO, LE, and RO+RE are then extracted by simply summing the five CCD rows with the highest flux.
(The spectrograph is set up carefully prior to each observing run so that all three spectra are parallel to the CCD rows.)
RO+RE represents the right circular polarization, while the left circular polarization can be obtained by combining LO and LE into a single spectrum.
Dark current subtraction is not necessary because our exposures are short and the CCD dark current is quite low.
Flat fielding is also unnecessary since both of the RO+RE and LO+LE spectra are exposed on the same pixels of the detector and subtraction of one from the other in the next processing step removes any pixel-to-pixel sensitivity variations.  Flat fields are normally acquired each run for possible use in the analysis of the combined intensity spectrum.

The magnetic shift observed in a single spectral line is measured by performing the Fourier cross-correlation of
the LO+LE and RO+RE spectra in a spectral window centered on the desired line, H$\beta$ in the case of \hdnum.
The magnetic shift is translated into a longitudinal magnetic field according to the line's magnetic sensitivity or the Land\'{e} factor, $g$ \citep[e.g., see][]{landstreet92}.
With $g=1$ for H$\beta$ the conversion factor between the magnetic shift observed in the DAO spectra and the longitudinal field strength is 6.8\,kG/pixel.
Individual longitudinal field measurements are then averaged and the error bar is calculated from the scatter of points.

A summary of the resulting magnetic field measurements for \hdnum\/ is given in Table \ref{magnetic_data} and includes the observation date, the observed field strength and error, and the rotation phase calculated as discussed below.
On most nights when \hdnum\/ was observed, bright, established magnetic Ap/Bp stars as well as bright non-magnetic stars were observed to confirm that the polarimeter was functioning properly.
More details of the performance of the polarimeter, including observations of these standards are described in Monin et al.\ (2011, in preparation).

\subsection{Photometry}

A total of 112 Hipparcos photometric measurements were retrieved from the Hipparcos archive \citep{esa97}.  We have not reprocessed these data in any way but simply use them below to determine the relative phasing between magnetic field, spectroscopic and photometric variations.

\section{Period Determination}

Variations in the H$\alpha$ profile of \hdnum\ became apparent after a few nights of observation with the DAO 1.8\,m telescope in 2005 August.  
At certain times weak emission features appear at velocities up to approximately $\pm450$\,km\,s$^{-1}$ from the line center while
at other times a traveling absorption feature is evident in the core of the line.  
The variability is very similar, although far less pronounced, to that seen in the helium-strong star $\sigma$\,Ori\,E and is what prompted us to search for a magnetic field in \hdnum.
An examination of Table \ref{magnetic_data} confirms that the star is obviously a strongly magnetic Bp star with a longitudinal field varying between approximately +2\,kG and -2\,kG.
We might then expect that the period of the magnetic field and spectral variations will measure
the rotation period of the star.

\cite{catanzaro99} obtained the first apparently robust period determination of $1.^{\!\!\rm{d}}58175$ for \hdnum.   A period of $1.^{\!\!\rm{d}}58193 \pm 0.00010$ is obtained from the Hipparcos data alone \citep{esa97,vanleeuwen07} and our spectroscopic and polarimetric observations are consistent with such a value.  However since our data have been obtained over several epochs we have been able to refine the rotation period of the star by combining our polarimetric and spectroscopic observations.

First, since the majority of peculiar magnetic stars have large-scale magnetic field geometries dominated by dipolar components and hence are expected to have approximately sinusoidal magnetic field curves, we have made a fit of our magnetic field observations to the equation $B_e=B_0+B_1\sin{2\pi(\phi-\phi_0)}$.
A period search between $0.^{\!\!\rm{d}}1$ and 100$^{\rm{d}}$
for all of our DAO magnetic field observations yields a single acceptable period of $1.^{\!\!\rm{d}}58215 \pm 0.^{\!\!\rm{d}}00070$, but with a relatively high $\chi^2/n$ of $2.29$.
However if the observation with the largest uncertainty is disregarded the $\chi^2/n$ is reduced to 1.87.
Our simple best-fit sinusoid gives values of $-40 \pm 63$\,G and  $-2032 \pm 14$\,G for $B_0$ and $B_1$ respectively.

Because of the single possible value of the rotation period provided by the magnetic field observations alone we have not conducted a detailed period analysis of our spectroscopic observations.
Instead we simply plotted the entire collection of DAO spectra in the region of the H$\alpha$ line phased on a range of periods consistent with the above magnetic period and its errors and with the continuum level of each spectrum offset by the value of its phase.
We then made qualitative estimates of the range of periods which gave an acceptable, smooth changes in the line profile variations with phase, especially when rapid variations are observed in the line core.
The significant time spanned by our spectroscopic observations led to a rotation period with a very large improvement in its uncertainty: $1.^{\!\!\rm{d}}5819840 \pm 0.^{\!\!\rm{d}}0000030$.
We have not made any attempt to search for evidence of a changing rotation period as has been seen recently in other members of the Bp stars \citep[e.g.][]{mikulasek08,townsend10} but \hdnum\/ will obviously be an interesting candidate for such a search in the future.

\begin{figure}
\centering
\includegraphics[angle=-90,scale=0.6]{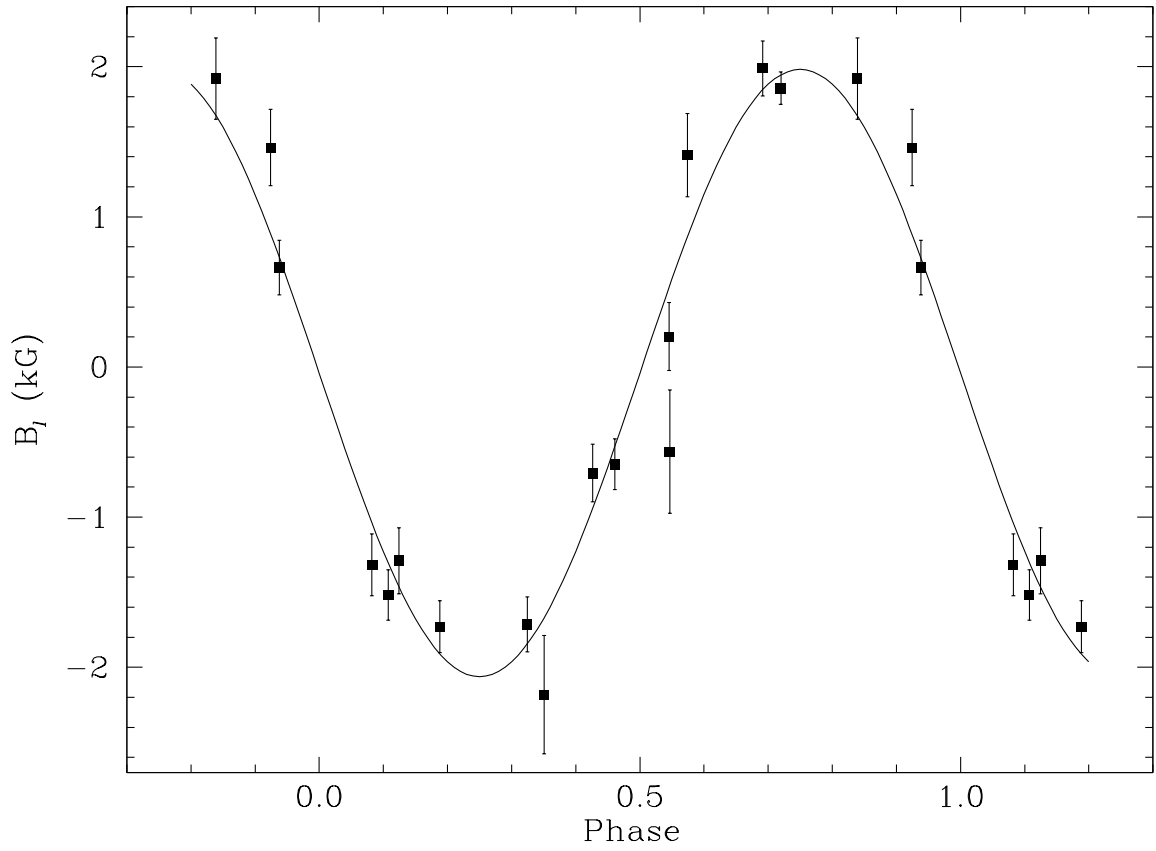}
\caption{Magnetic field variations of \hdnum\/ phased on the ephemeris provided in Equation \ref{ephem}.  The solid line is the best sinusoidal fit to the data and is used to define the magnetic geometry discussed in the text.\label{magfield}}
\end{figure}

In Figures \ref{magfield} through \ref{halpha} we show our magnetic field measurements, the Hipparcos photometry and the observed H$\alpha$  line variations all phased on the ephemeris

\begin{eqnarray}
\rm{JD}(B_e^0)=(2454496.6938 \pm 0.0017) +  \nonumber \\
(1.5819840\pm0.0000030) \cdot \rm{E}. \label{ephem}
\end{eqnarray}

\noindent The zero point was chosen to lie near the midpoint of the spectroscopic observations and has been set in such a way that $\phi = 0$ occurs when the best-fit field curve changes sign from positive to negative. 

\section{Spectrum Synthesis}

Published values of the temperature and surface gravity of \hdnum\ have a substantial range.
Str\"{o}mgren photometry and various calibration systems were used by \citet{molenda04} and \citet{catanzaro99} to find $T_{\rm eff} = 18\,071$ and $\log{g} = 4.42$ and $T_{\rm eff} = 18\,300$ and $\log{g} = 4.31$ respectively.  
Most recently, \citet{huang08} gave $T_{\rm eff} = 15\,538 \pm 150$\,K and $\log{g} = 3.727 \pm 0.024$ from spectrum modeling which included the possible effects of rapid rotation.
Because of these substantial difference and the availability of our high-S/N and high-resolution CFHT data we have performed our own spectral modelling to try to resolve the discrepancy in the previously derived fundamental parameters of \hdnum.

Our model spectra were generated with non-LTE model atmospheres produced with the TLUSTY program \citep{hubeny95} as input for
the spectrum synthesis program SYNSPEC (version 48 provided by Lanz, private communication) and the accompanying rotational convolution program Rotin3.  
We adopted atmospheres from the solar composition (BG) grid generated for B-type stars by \cite{lanz07}.
We used our CFHT ESPaDOnS spectra and the \ion{Si}{2} $\lambda\lambda$4128 and 4130 lines and \ion{Si}{3} $\lambda\lambda$4552, 4567, and 4574 lines to determine an approximate $T_{\rm eff}$ for \hdnum.

Because of concerns with continuum fitting of the ESPaDOnS data in regions containing broad hydrogen Balmer lines and helium lines we instead used our DAO H$\alpha$ observations to determine $\log{g}$ and confirm the $T_{\rm eff}$ derived from the relative strengths of the \ion{Si}{2}/\ion{Si}{3} lines.
Since weak emission contaminates the wings of the H$\alpha$ profile near rotation phases 0.25 and 0.75 we averaged all of the DAO 1.8\,m spectra obtained when there is no obvious H$\alpha$ emission in the line wings, i.e. between phases 0.95 and 0.05 as well as between 0.45 and 0.55.
We then generated synthetic spectra to compare with these average spectra, paying no attention to deviations between the model and observed spectra in the central 5\,\AA\/ of the line because of the obvious variable absorption feature that travels through the line core at these two phases.
In a similar way we averaged the DAO spectra obtained between phases 0.70 and 0.80, when no variable absorption appears to be present in the core, and compared this mean spectrum with the same model spectra to compare the consistency of the model with the observations in the line core.

Our adopted model has a $T_{\rm eff} = 16\,000$\,K and $\log{g} = 4.0$ and the resulting synthetic spectrum in the region of H$\alpha$ is shown compared to the averaged observations at phases 0.500 and 0.750 in Figure~\ref{halpha_model}.
In the line synthesis we adjusted the helium abundance to one-half solar and we adopted five times solar abundances for C, O, Si, S, and Fe.  
Such abundance enhancements are not atypical for a magnetic Bp star.
We note, however, that a few additional models we carried out by generating TLUSTY atmospheres with a stratified helium abundance and somewhat higher temperatures can also provide a reasonable good fit to the observed hydrogen and helium line profiles.
We have not, however, pursued this investigation in any detail since the helium lines, and hence helium abundance, are obviously variable in strength (see below) but the S/N, resolution and wavelength coverage of our DAO spectra are insufficient to permit a more detailed analysis of the possible non-uniform horizontal and vertical abundance distributions of helium at this time.
A much more extensive set of phase-resolved ESPaDOnS spectra, perhaps obtained in spectropolarimetric mode, would be very useful for a more comprehensive study of this star.

Our modelling of the metal lines of \hdnum\/ give a $v\sin{i}$ of $105\pm10$\,km\,s$^{-1}$ and a radial velocity of $-16\pm2$\,km\,s$^{-1}$, the latter consistent with published values \citep{bsc}.
\cite{huang08} have recently measured $v\sin{i} = 119\pm13$\,km\,s$^{-1}$ for the star using \ion{He}{1} lines as well as \ion{Mg}{2} $\lambda4481$.
Their result included the effects of Roche geometry and gravity darkening which our analysis ignores.
This might contribute to their somewhat higher $v\sin{i}$ although our values agree within their formal errors.
The $v\sin{i}$ value of 65\, km\,s$^{-1}$ measured by \citet{abt02} is not acceptable.

\section{Magnetic Field Geometry} 

The Hipparcos parallax for \hdnum\ is $\pi = 3.42\pm0.30$\,mas \citep{vanleeuwen07} so the star's distance is $r = 292^{+28}_{-24}$\,pc.
Adopting magnitudes from Table~\ref{properties}, the reddening-independent index
$Q = (U-B) - 0.72(B-V)$,
and $(B-V)_0 = Q/3$, we find $E(B-V) = 0.051$ and hence $A_V = 3.1 \times E(B-V) = 0.157$.
The absolute magnitude of \hdnum\ is then $M_V = -1.08^{+0.18}_{-0.20}$.

Using $T_{\rm eff} = 16\,000 \pm 1000$\,K from our spectrum synthesis, we derive $BC = -1.38\pm0.16$ from the calibration of \cite{balona94}.
An $M_{bol}(\odot)$ value of $+4.75$ \citep{allen76} yields $log{(L/L_{\odot})} = 2.88\pm0.14$.
We then find possible values of $R=3.61^{+0.56}_{-0.45}\,R_{\odot}$ if we adopt $T_{\rm eff\odot} = 5780$\,K.

Since we know the rotation period of the star we can determine the inclination of the rotation axis, $i$, using
\begin{equation}
\sin{i} = \frac{Pv\sin{i}}{50.6R}
\end{equation}
with the period, $P$, expressed in days, $v\sin{i}$ in km\,s$^{-1}$, and the radius, $R$, in solar units.
Adopting $v\sin{i} = 105\pm10$\,km\,s$^{-1}$ from our spectrum synthesis described above we find a lower limit to the inclination of
$i\ge 52\degr$.
This limit is only very weakly sensitive to the effective temperature of the star.

For a dipolar magnetic field with an obliquity, $\beta$, of the magnetic axis to the rotation axis, the angles $i$ and $\beta$ are related by 
\begin{equation}
\tan{\beta} = \left(\frac{1-r}{1+r}\right)\cot{i}
\end{equation}
where $r = (|B_0|-B_1)/(|B_0|+B_1) = -0.96^{+0.6}_{-0.2}$ according to our best-fit sinusoid.
For such values of $r$ near -1 either $i$ or $\beta$ must be close to 90\degr.
In the next section because of geometrical considerations we argue that the inclination must be close to $90\degr$.
If we adopt $i=85$\degr\/ then $\beta = 77$\degr.  We need information about the surface field strength to further constrain $\beta$ or $i$ but we can derive a lower limit to the polar surface field strength from the relation 
\begin{equation}
B_{\rm d} = B^{\rm max}_l\left(\frac{15+u}{20(3-u)}(\cos{\beta}\cos{i}+\sin{\beta}\sin{i})\right)^{-1}
\end{equation}
where $u$ is the limb darkening coefficient \citep{preston67}.
This gives $B_{\rm d} \geq 7$\,kG and an equatorial field strength $B_{\rm eq} \geq 2.7$\,kG for a value of $u=0.4$.

Table~\ref{properties} provides a summary of the observed and derived properties of \hdnum.

\section{Discussion}

\begin{figure}
\centering
\includegraphics[angle=0,scale=0.75]{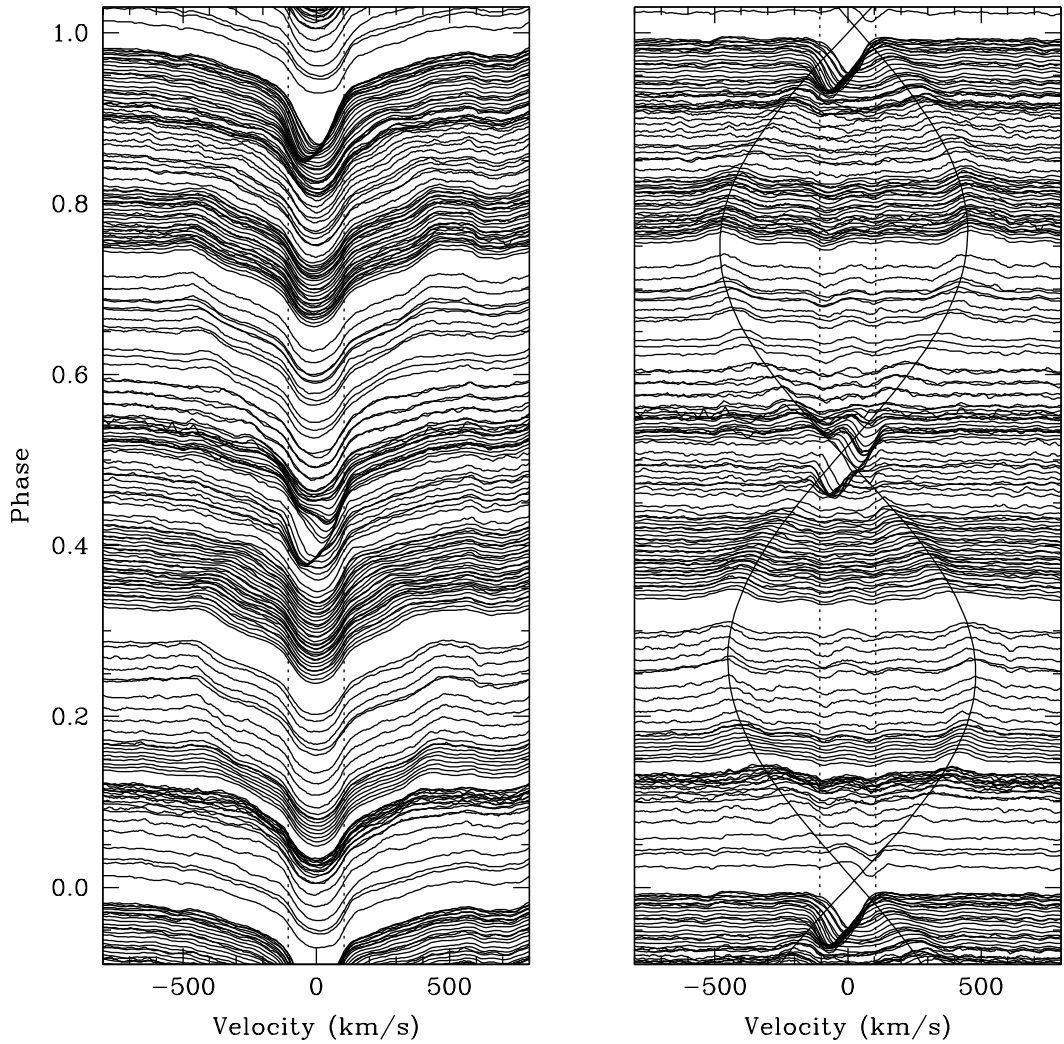}
\caption{Observations of the H$\alpha$ profile variations of \hdnum.  The left panel shows the continuum normalized intensity spectra.  Data are plotted in such a way that the continuum of each spectrum is plotted at a height that corresponds to the phase determined from the ephemeris provided in Equation \ref{ephem}.  Dashed vertical lines are drawn to show the $v\sin{i}$ of the star.  The right panel shows the residual spectra after division by a model spectrum with $T_{eff} = 16\,000$ and $\log{g}=4.0$ (see text).  Note the weak emission features which are most obvious at $\pm450\,$km\,s$^{-1}$ from line center near phases 0.25 and 0.75 as well as the absorption feature that moves through the line center near phases 0.0 and 0.5.  Curves in the right panel help show the motion of the emission features as discussed in the text. \label{halpha}}
\end{figure}

\begin{figure}
\centering
\includegraphics[angle=-90,scale=.6]{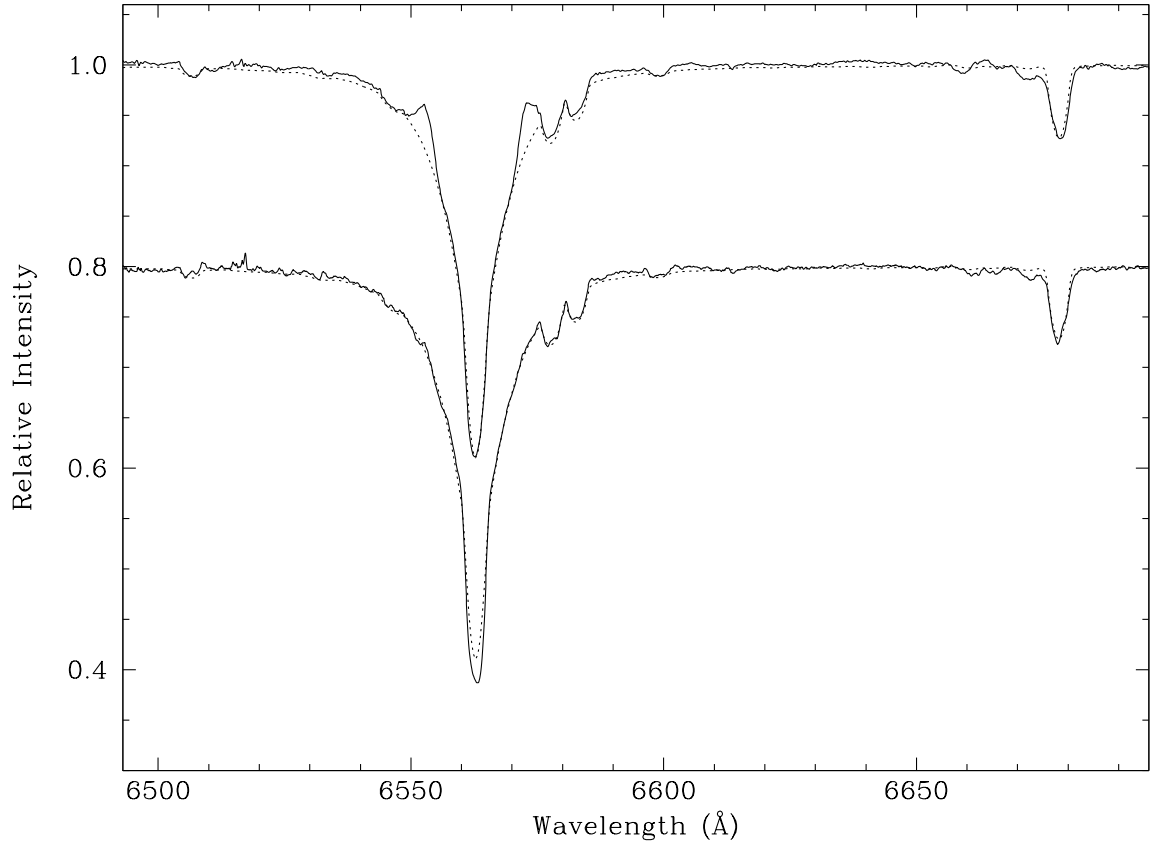}
\caption{The dashed lines show a model TLUSTY/SYNSPEC spectrum with $T_{\rm eff} = 16\,000$\,K, $\log{g}=4.0$, one-half solar helium abundance and five times solar carbon abundance compared to an average of the observed spectrum of \hdnum\/ near phase 0.5 (top) when no shell absorption is present in the observed H$\alpha$ profile, and phase 0.0 (bottom) when emission does not contaminate the wings of the line.  This model spectrum is used to produce the residual spectra shown in Figure~\ref{halpha} and Figure~\ref{grayscale}.\label{halpha_model}}
\end{figure}

We have already briefly discussed the observed periodicity in the variability of the H$\alpha$ line of \hdnum\/ shown in the left panel of Figure~\ref{halpha}.
However the variations are much more obvious if we remove the photospheric component of the line profile and we have done this by dividing each observation by the model profile discussed above and shown in Figure~\ref{halpha_model}.  
The results are shown in the right panel of Figure~\ref{halpha} and a grayscale representation is shown in Figure~\ref{grayscale}.

\begin{figure}
\centering
\includegraphics[angle=-90,scale=0.6]{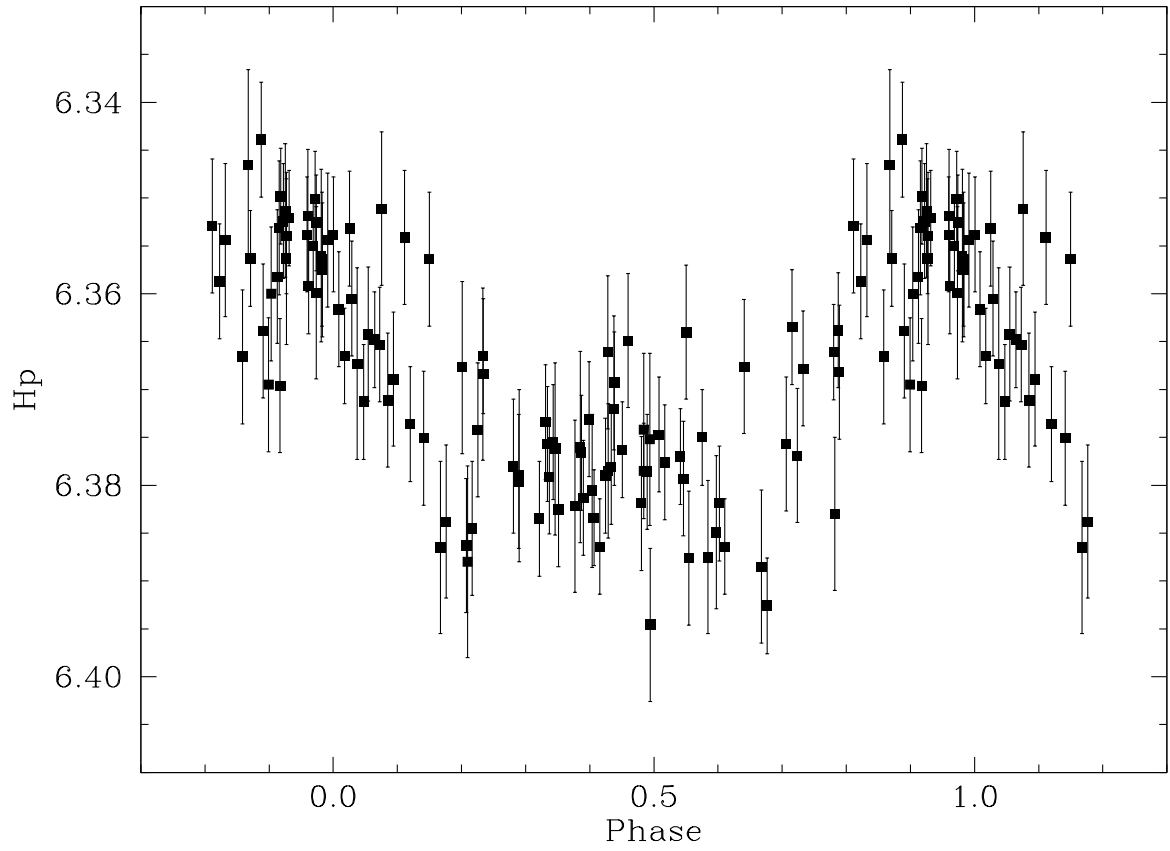}
\caption{Hipparcos photometry of \hdnum\/ phased on the ephemeris provided in Equation~\ref{ephem}.\label{hipparcos}}
\end{figure}

There are emission and absorption features that show variability on a period that also results in well-defined magnetic field 
(Figure~\ref{magfield}) and photometric (Figure~\ref{hipparcos}) variations.
The material that gives rise to this emission must be circumstellar since it displays velocities in excess of 400\,km\,s$^{-1}$, much larger than the measured $v\sin{i}$ of the star.
If we assume that the period corresponds to the rotation period of the star then this implies that the circumstellar gas corotates with the stellar surface; i.e. the magnetic field of the star must force the material into rigid-body rotation.
We can use this fact in Figure~\ref{grayscale} to add a second scale on the abscissa that corresponds to the radial distance from the center of the star.

Near rotation phases 0.0 and 0.5, corresponding to the zero-crossings of the magnetic field curve, the emission disappears for the most part and is replaced by absorption features that move from short to long wavelengths through the H$\alpha$ line core.
This presumably arises when the circumstellar material also crosses the line of sight to the observer since these features are only seen at velocities less than approximately 110\,km\,s$^{-1}$.
This also provides an independent (and consistent) measurement of the $v\sin{i}$ of the star.

The H$\alpha$ spectrum variability and magnetic field observations of  \hdnum\/ agree with the picture of magnetically confined winds and rigidly-rotating magnetospheres for magnetic Bp stars \citep{babel97a, babel97b,uddoula02,townsend05a, townsend07}.
This scenario proposes that radiatively driven winds are likely to occur in B stars with strong, large scale magnetic fields but with a mass-loss topology strongly affected by the structure of the magnetic field at the star's surface. 
Above the stellar surface, an approximately dipolar field will force the wind component from the two hemispheres towards the magnetic equator, where the wind from the two hemispheres will collide and lead to a strong shock and the formation of a magnetosphere.  The magnetosphere consists of an extended, hot post-shock region as well as a dense cooling disk formed near the magnetic equator and corotating with the star several stellar radii above its surface.
X-rays are emitted from the shock region and electrons can be accelerated to energies required for radio emission in the GHz band.
Such a model may also explain the variable satellite UV resonance lines seen such objects \citep{shore87,shore90}.

\hdnum\/ has not been detected in the X-ray domain but the ROSAT all-sky survey upper limit of $\log{(L_X/L_{\rm bol})}=-6.58$ is  more than two times the ROSAT-PSPC pointed observation detection of $\log{(L_X/L_{\rm bol})}=-6.90$ for the A0p star IQ Aur \citep{babel97a}.
Unfortunately IUE data are not available for \hdnum\/ so we can not say anything about variability of the satellite UV lines.
As already mentioned in the introduction, it was the detection of radio emission from the star \citep[Drake, private communication;][]{drake06} that first drew our attention to it.

\cite{uddoula02} define the wind magnetic confinement parameter as 
\begin{equation}
\eta_* \equiv \frac{B^2_{\rm eq} R^2_*}{\dot{M}v_{\infty}}.
\end{equation}
For parameters provided in Table~\ref{properties},  and estimated values of $\dot{M}=10^{-10}$\,M$_{\odot}$\,yr$^{-1}$ and $v_{\infty}=800$\,km\,s$^{-1}$ for a star like \hdnum\/ \citep{babel97a}, $\eta_* \approx 2.3\times10^5$ and the weak wind of \hdnum\/ can be expected to be very strongly confined.
It is therefore not surprising that the star displays the same phenomena observed in other hot, strongly magnetic Bp stars such as $\sigma$\,Ori\,E \citep{townsend05b,smith07,oksala11}, $\delta$\,Ori\,C \citep{leone10}, and 36\,Lyn \citep{wade06,smith06} also known to have magnetically confined winds.

Examination of Figs.~\ref{halpha} and \ref{grayscale} leads to the following observations regarding the circumstellar environment of \hdnum:
\begin{enumerate}
\item The H$\alpha$ emission comes from two distinct regions which manifest themselves in Figs.~\ref{halpha} and \ref{grayscale}  as a double S-wave.
If we interpret these observations within the framework of current magnetically confined wind and magnetosphere models we can assume that these regions are located in the magnetic equatorial plane.

\item The two emission regions are almost, but not quite, diametrically opposite each other, and rotate on similar circular orbits around the star.  
In the right hand panel of Figure~\ref{halpha} we have drawn sinusoids through the peaks of each of the emission regions.  
The curves have amplitudes of  450 and $480$\,km\,s$^{-1}$ (corresponding to distances of 4.3 and 4.6\,R$_*$ from the center of the star) but are offset from each other by 6\degr\/ or 0.017 in phase.
The sector that is moving in a redward direction at $\phi=0.0$ is slightly further away from the star and appears to cross in front of the star at almost precisely $\phi=0.0$.  
The center of the other sector trails slightly, moving behind the star about 0.017 cycles later, and the peak emission is also marginally brighter.

\item Both sectors have radial extents of about 1.8\,R$_*$.  Near phases 0.25 and 0.75 the outer edge of the emission is at about 5.5\,R$_*$ for both sectors and the inner edge is at about 3.7\,$R_*$.
For both the emission appears to decline somewhat more rapidly towards the stellar surface than it does towards the outer edge of each sector.

\item The inclination of the rotation axis of the star, $i$, must be close to 90\degr\/ given the substantial distance of the emitting sectors from the star and the fact that very little central emission is seen in the S-wave curves as they cross zero velocity.   At these phases one sector must be almost completed occulted by the star while the other sector passes in front of the stellar disk.

\item The azimuthal extent of the sectors can be determined to a first approximation by timing the duration of the shell episodes.
Unfortunately phase coverage near $\phi=0.0$ is not very complete so we concentrate mainly on the event at $\phi=0.5$.
The first hint of a shell feature in the blue wing of H$\alpha$ occurs at $\phi=0.453$ while the feature disappears from the red wing at approximately $\phi=0.551$, or a duration of 0.098 in phase.
Converting this time into a measurement of the size of the sector in the plane of rotation obviously depends on the geometry and other physical characteristics of the circumstellar clouds.  However, to first order this is the time it takes the circumstellar sector at a given distance from the star to move a distance of about $2\,R_*+X$ where $X$ is the azimuthal extent of the sector at that distance.
We then easily determine that $X$ can be no larger than $0.3\,R_*$ at a distance of $2.7\,R_*$ from the star's surface.
The circumstellar material could extend to $1.4\,R_*$ at a distance of $4.5\,R_*$ from the star from geometrical considerations only but this does not appear to be likely given how narrow the shell features are at phases 0.0 and 0.5 (see the final point below).
At the point of peak emission the circumstellar emission can extend no further than $\approx 0.6\,R_*$ in the azimuthal direction and this is likely a more representative measure.
The $\phi=0.0$ event looks very similar which suggests that the other sector has a similar size.

\item Estimating the meridional extent of the sectors is difficult but it does not appear to be large. Since emission from one of the sectors disappears quite abruptly at about $\phi=0.45$ when it is occulted by the star this suggests that this sector is extended no more than $2\,R_*$ in this direction if $i \approx 90\degr$.
The other sector may be slightly more extended than $2\,R_*$ since it is somewhat brighter (and if other physical properties for the two sectors are the same) and its emission does not disappear completely between $\phi=0.95$ and 0.05 as it moves behind the star.
The shell absorption produced by this sector at $\phi=0.5$ is also not as deep as the event at $\phi=0.0$ which might be caused by filling in by emission if a small portion of this sector extends beyond the limb of the star.

\item The right panel of Figure~\ref{halpha} shows that at phases 0.0 and 0.5 the shell features are narrower than the projected $v\sin{i}$ of the star.
Since we have argued that the emitting regions are most extended radially and along the magnetic equator, this implies a large magnetic obliquity, $\beta$, consistent with our discussion of the magnetic field geometry above.
This also suggests that the azimuthal extents of the emission sectors are likely smaller than the maximum value provided by geometrical arguments alone.

\end{enumerate}

\begin{figure}
\centering
\includegraphics[angle=-90,scale=0.6]{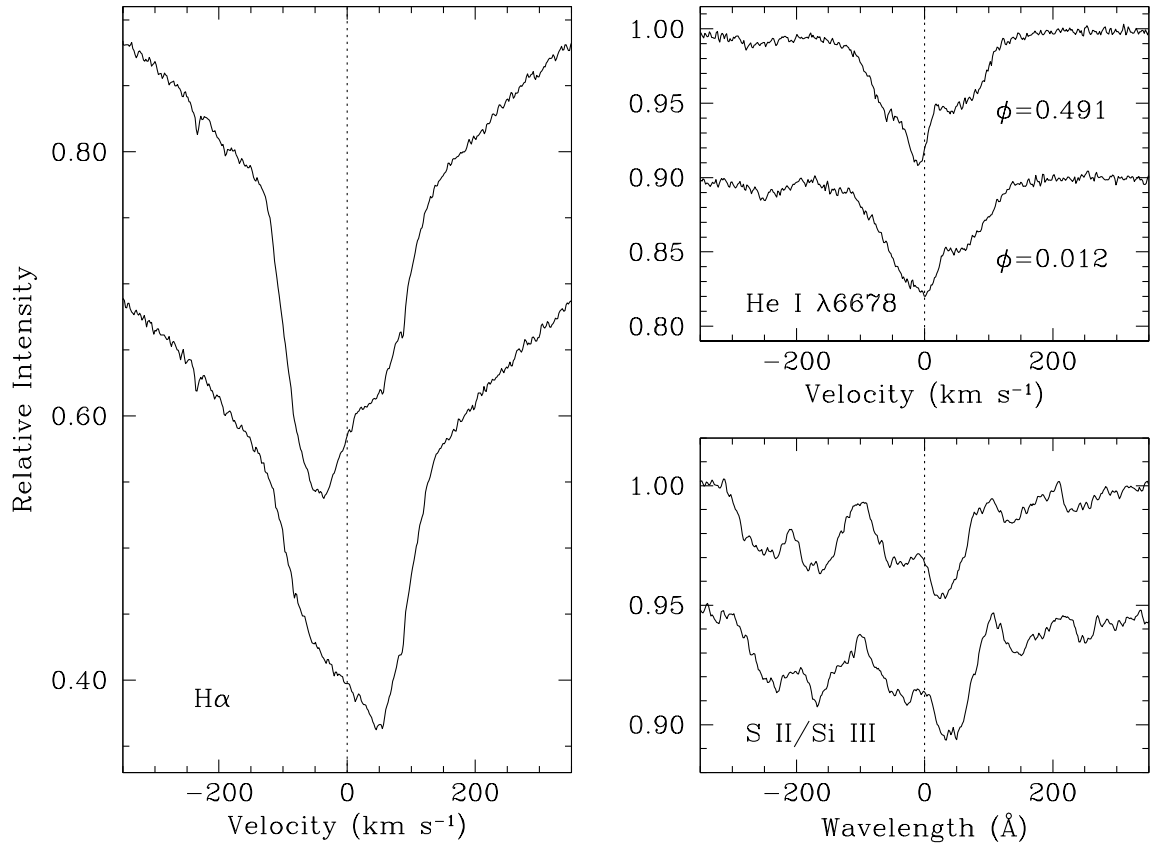}
\caption{Portions of CFHT ESPaDOnS spectra of \hdnum\/ in three different wavelength regions and at two different phases are shown to illustrate variability in helium and metal lines compared to the shell absorption features in the core of the H$\alpha$ profile.  
Both spectra were obtained near alternate magnetic zero crossings with phases indicated in the top right panel.\label{detail}}
\end{figure}

Hydrogen is not the only element with line variability in \hdnum.
While the resolution and S/N of the DAO spectra are not high enough to show this very clearly, the two CFHT ESPaDOnS spectra were serendipitously obtained near the two zero crossings in the magnetic field curve at phases 0.012 and 0.491.
Just as H$\alpha$ displays prominent shell absorption at these phases, the \ion{He}{1} lines of the star also show pronounced differences.
In Figure~\ref{detail} we illustrate spectra of the core of the H$\alpha$ line, the \ion{He}{1} $\lambda$6678 line as well as \ion{Si}{3} and \ion{S}{2} lines at both phases.
In both spectra the \ion{He}{1} line profile consists of a relatively narrow absorption feature near the line center superimposed on an underlying broad profile.
These narrow \ion{He}{1} components cannot be physically associated with the circumstellar material made apparent by the shell features in the core of the H$\alpha$ line.  
Not only do the \ion{He}{1} features fall at substantially lower velocities than those of H$\alpha$, but it is also unrealistic to expect any cool circumstellar helium to be excited to a high enough excitation level to produce circumstellar shell features.
Instead, the \ion{He}{1} profiles likely indicate that \hdnum\/ has an enriched helium content along its magnetic equator relative to the rest of its surface.

In contrast to helium, the \ion{Si}{3} $\lambda$4552 line also shown in Figure~\ref{detail} displays two components on either side of line center. This may suggest that silicon has an enhanced abundance at both magnetic poles of the star.
The nearby \ion{S}{2} feature as well as lines of other elements show similar behavior.
Clearly it would be of interest to obtain additional ESPaDOnS spectra of \hdnum\/ to investigate its abundance geometry in detail.

The photometric variations of magnetic Bp stars such as \hdnum\/ (Figure~\ref{hipparcos}) are thought to arise because of their inhomogeneous surface abundances and, as a result, flux redistribution caused by phase-dependent absorption \citep{krticka07}.
\hdnum\/ is brightest at phases 0.95 and 0.45, just before the magnetic equator crosses the line of sight to the observer and when, from our discussion above, the helium lines are strongest.
We see no evidence for photometric ``eclipses'' caused by the transit of the circumstellar material as are seen in $\sigma$\,Ori\,E, but more precise photometry would be of interest.

\section{Conclusion}

A strong, reversing, approximately dipolar magnetic field and weak, variable H$\alpha$ emission and shell absorption have been discovered in the helium-weak star \hdnum.  The magnetic field, spectrum and archival Hipparcos photometry all show well-defined variability on a period of $1.^{\!\!\rm{d}}5819840 \pm 0.^{\!\!\rm{d}}0000030$.
From non-LTE models of the star's spectrum, the magnetic field curve, and the derived geometry of the circumstellar material that gives rise to the observed H$\alpha$ emission, we determine that the inclination, $i$, of the rotation axis of the star must be close to 90\degr.   If we adopt a value of $i=85$\degr\/ then the magnetic obliquity, $\beta$, is about 77\degr.

The star's circumstellar material consists of two very similar emitting volumes located at the intersections of its magnetic and rotation equators.
The strong magnetic field of \hdnum\/ forces this material into corotation at a distance of approximately $3.5\,R_*$ above its surface.
From basic geometrical considerations we estimate radial sizes of $1.8\,R_*$ and meridional extents of $2\,R_*$ or less for both circumstellar sectors.
Their azimuthal extent, perpendicular to the magnetic equator, is approximately $0.6\,R_*$.

The observed spectral variations of \hdnum\/ show many similarities to those of other magnetic helium-peculiar stars such as $\sigma$\,Ori\,E, $\delta$\,Ori\,C and 36\,Lyn.   
A few things, however, make \hdnum\/ a particularly intriguing object.
First, with a $T_{\rm eff}$ of about 16\,000\,K, it is the coolest such object to show variable H$\alpha$ emission in addition to shell absorption events.
Secondly, previous detailed studies of the magnetic field and circumstellar geometries of $\sigma$\,Ori\,E \citep{short94,groote97,townsend05b,oksala11} and $\delta$\,Ori\,C \citep{leone10} have shown that both of these stars have quite complex circumstellar and magnetic field geometries.
In addition, the former object has intermediate values for $i$ and $\beta$ while the latter has a small $i$ and suffers from the complication of being a double-lined spectroscopic binary.
\hdnum, on the other hand, has large values of $i$ and $\beta$ and apparently quite simple magnetic field and circumstellar geometries.
This should make it an interesting object for future modelling with tools such as the Rigidly Rotating Magnetosphere (RRM) model \citep{townsend05a}.

Finally, the precision with which we have been able to determine the rotation period of \hdnum\/ makes it a very promising candidate to search for evidence of magnetic breaking similar to that reported by \cite{mikulasek08} for HD\,37776 and \cite{townsend10} for $\sigma$\,Ori\,E.
Hints of pronounced spectrum variability of other elements including helium, silicon and sulphur will also make it an interesting object for detailed high-resolution, magnetic Doppler imaging.
Such a study will hopefully provide an improved understanding of the interaction between a star's surface magnetic field geometry, its radiatively driven wind, diffusion processes in its photosphere, and energetic processes taking place in its magnetosphere as described by current and future improved models of rigidly rotating magnetospheres.

\acknowledgments

\begin{figure}
\centering
\includegraphics[angle=0,scale=1.2]{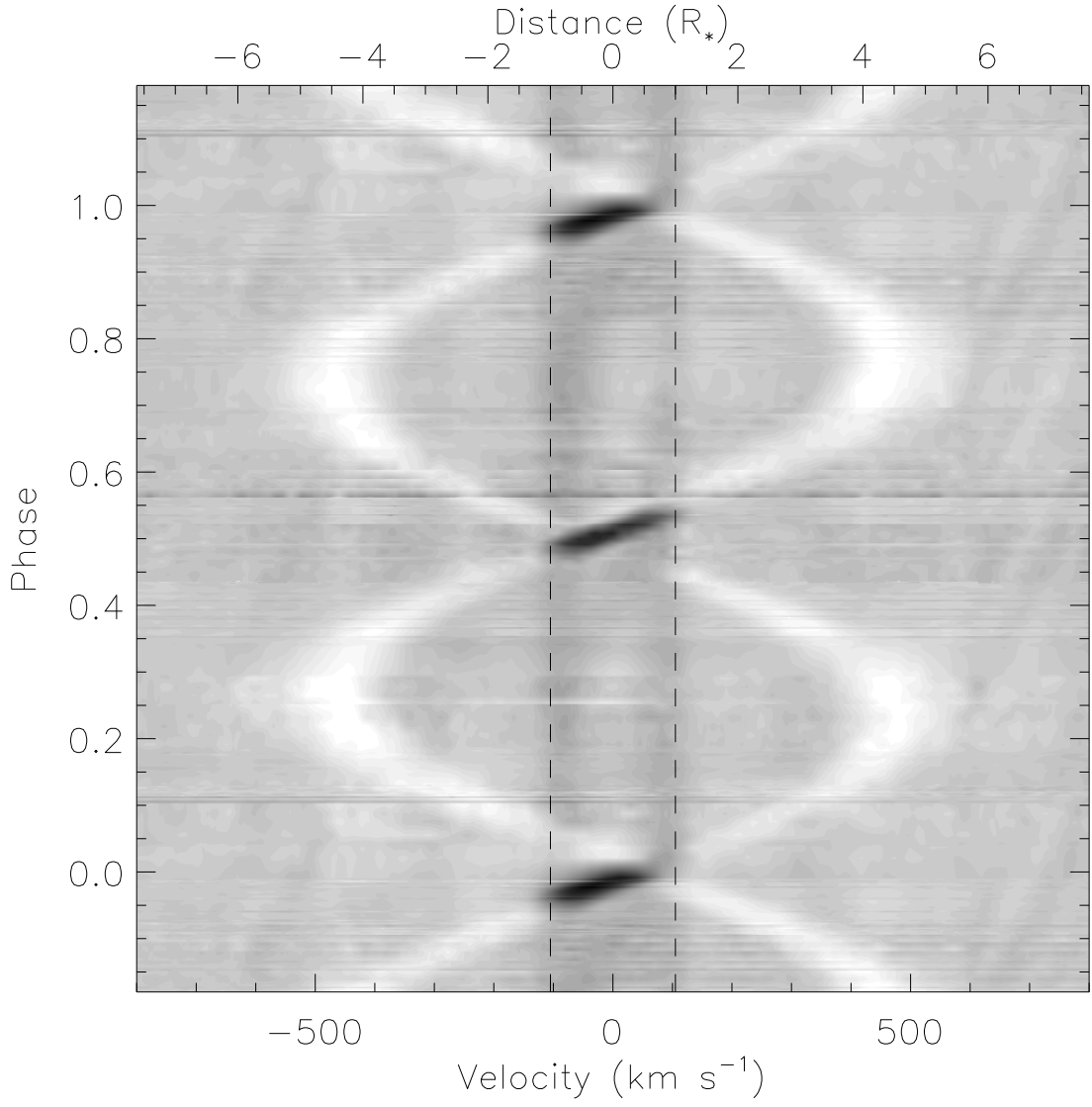}
\caption{A grayscale representation of the residual H$\alpha$ profiles of \hdnum\/ shown in the right panel of Figure~\ref{halpha}.   The dashed vertical lines denote the $v\sin{i}$ of the star.  The distance scale shown at the top of the plot is derived from the velocity scale by assuming that the circumstellar gas is constrained by the strong magnetic field to rotate rigidly about the star.\label{grayscale}}
\end{figure}

The authors would like to thank Dr. Franco Leone for an IDL procedure which, with some modifications, was used to produce Figure~\ref{grayscale}.

{\it Facilities:} \facility{DAO:1.22m (McKellar Spectrograph)}, \facility{DAO:1.85m (Cassegrain spectrograph, spectropolarimeter)}, \facility{CFHT (ESPaDOnS)}




\clearpage

\clearpage

\clearpage

\clearpage

\clearpage

\clearpage

\clearpage

\clearpage

\begin{deluxetable}{rcccrcccrcc}
\tabletypesize{\scriptsize}
\tablecaption{Journal of spectra of \hdnum.\label{spectra}}
\tablewidth{0pt}
\tablehead{
\colhead{Obs. \#} & \colhead{HJD} & \colhead{$\phi$} & ~~~& \colhead{Obs. \#} & \colhead{HJD} & \colhead{$\phi$}  & ~~~ & \colhead{Obs. \#} & \colhead{HJD} & \colhead{$\phi$} \\
& (-245\,0000)  & & & & (-245\,0000) & & & & (-245\,0000)
}
\startdata
\sidehead{DAO 1.2\,m data}
2000\_010461 & 1770.7107 & 0.857 & & 2000\_010462 & 1770.7252 & 0.867 & & 2000\_010529 & 1771.7133 & 0.491\\ 
2005\_015601 & 3621.7198 & 0.913 & & 2005\_015603 & 3621.7413 & 0.927 & & 2005\_015605 & 3621.7628 & 0.940\\ 
2005\_018890 & 3641.6863 & 0.534 & & 2005\_018892 & 3641.7078 & 0.548 & & 2005\_018894 & 3641.7294 & 0.561\\ 
2005\_018896 & 3641.7509 & 0.575 & & 2005\_018898 & 3641.7724 & 0.589 & & 2005\_018900 & 3641.7940 & 0.602\\ 
2005\_019114 & 3645.7049 & 0.074 & & 2005\_019115 & 3645.7257 & 0.088 & & 2005\_019116 & 3645.7472 & 0.101\\ 
2005\_019117 & 3645.7681 & 0.114 & & 2006\_001891 & 3786.0311 & 0.777 & & 2006\_019493 & 4013.6531 & 0.661\\ 
2006\_019495 & 4013.6748 & 0.675 & & 2006\_019498 & 4013.7098 & 0.697 & & 2006\_019500 & 4013.7314 & 0.711\\ 
2006\_019502 & 4013.7531 & 0.724 & & 2006\_019654 & 4015.7090 & 0.961 & & 2006\_019657 & 4015.7353 & 0.977\\ 
2006\_019817 & 4017.6333 & 0.177 & & 2006\_019819 & 4017.6550 & 0.191 & & 2006\_019821 & 4017.6767 & 0.204\\ 
2006\_019823 & 4017.6983 & 0.218 & & 2006\_019825 & 4017.7200 & 0.232 & & 2008\_009030 & 4645.7974 & 0.251\\ 
2008\_009031 & 4645.8185 & 0.264 & & 2008\_009032 & 4645.8396 & 0.277 & & 2008\_009033 & 4645.8607 & 0.291\\ 
2008\_009124 & 4646.7978 & 0.883 & & 2008\_009125 & 4646.8190 & 0.896 & & 2008\_009126 & 4646.8401 & 0.910\\ 
2008\_009127 & 4646.8612 & 0.923 & & 2010\_011433 & 5381.7608 & 0.466 & & 2010\_011435 & 5381.7826 & 0.480\\ 
2010\_011437 & 5381.8043 & 0.493 & & 2010\_011439 & 5381.8260 & 0.507 & & 2010\_011441 & 5381.8478 & 0.521\\ 
2010\_011443 & 5381.8695 & 0.535 & & 2010\_011445 & 5381.8912 & 0.548 & & 2010\_011447 & 5381.9130 & 0.562\\ 
2010\_011449 & 5381.9347 & 0.576 & & 2010\_011451 & 5381.9564 & 0.590 & & 2010\_011453 & 5381.9782 & 0.603\\ 
2010\_011605 & 5384.7451 & 0.352 & & 2010\_011606 & 5384.7662 & 0.366 & & 2010\_011607 & 5384.7873 & 0.379\\ 
2010\_011608 & 5384.8084 & 0.392 & & 2010\_011609 & 5384.8295 & 0.406 & & 2010\_011610 & 5384.8506 & 0.419\\ 
2010\_011611 & 5384.8717 & 0.432 & & 2010\_011612 & 5384.8928 & 0.446 & & 2010\_011613 & 5384.9139 & 0.459\\ 
2010\_011614 & 5384.9350 & 0.472 & & 2010\_011615 & 5384.9561 & 0.486 & & 2010\_011616 & 5384.9772 & 0.499\\ 
2010\_012097 & 5391.7372 & 0.772 & & 2010\_012098 & 5391.7583 & 0.786 & & 2010\_012099 & 5391.7793 & 0.799\\ 
2010\_012100 & 5391.8004 & 0.812 & & 2010\_012101 & 5391.8215 & 0.826 & & 2010\_012102 & 5391.8425 & 0.839\\ 
2010\_012103 & 5391.8636 & 0.852 & & 2010\_012104 & 5391.8847 & 0.865 & & 2010\_012105 & 5391.9057 & 0.879\\ 
2010\_012106 & 5391.9268 & 0.892 & & 2010\_012107 & 5391.9479 & 0.905 & & 2010\_012108 & 5391.9689 & 0.919\\ 
\sidehead{DAO 1.8\,m data}
1999\_001392 & 1249.0469 & 0.105 & & 1999\_001393 & 1249.0586 & 0.112 & & 2005\_008230 & 3603.8585 & 0.623\\ 
2005\_008306 & 3605.8012 & 0.851 & & 2005\_008331 & 3606.7819 & 0.471 & & 2005\_008469 & 3607.8243 & 0.129\\ 
2005\_008596 & 3608.8273 & 0.763 & & 2005\_008646 & 3609.8068 & 0.383 & & 2005\_008709 & 3610.7687 & 0.991\\ 
2005\_008775 & 3611.7829 & 0.632 & & 2005\_008846 & 3614.7895 & 0.532 & & 2005\_012635 & 3718.5453 & 0.118\\ 
2005\_012636 & 3718.5494 & 0.121 & & 2005\_012637 & 3718.5529 & 0.123 & & 2005\_012639 & 3718.5571 & 0.126\\ 
2005\_012640 & 3718.5606 & 0.128 & & 2005\_012641 & 3718.5640 & 0.130 & & 2006\_000316 & 3777.0715 & 0.114\\ 
2006\_000317 & 3777.0809 & 0.120 & & 2006\_000517 & 3778.1014 & 0.765 & & 2006\_000878 & 3782.1060 & 0.296\\ 
2006\_001185 & 3783.0706 & 0.906 & & 2006\_001186 & 3783.0776 & 0.910 & & 2006\_001187 & 3783.0845 & 0.915\\ 
2006\_004572 & 3863.9876 & 0.055 & & 2006\_004717 & 3864.9853 & 0.685 & & 2006\_007422 & 3932.9775 & 0.665\\ 
2006\_009439 & 3952.8913 & 0.252 & & 2006\_009499 & 3953.7752 & 0.811 & & 2006\_009500 & 3953.7843 & 0.817\\ 
2006\_009615 & 3955.8285 & 0.109 & & 2006\_009777 & 3958.7853 & 0.978 & & 2006\_009778 & 3958.7943 & 0.984\\ 
2006\_009779 & 3958.7999 & 0.987 & & 2006\_012752 & 3984.7127 & 0.367 & & 2006\_012896 & 3985.7465 & 0.021\\ 
2006\_012988 & 3986.7235 & 0.638 & & 2006\_013559 & 4012.6652 & 0.037 & & 2006\_013561 & 4012.6741 & 0.042\\ 
2007\_014554 & 4426.5948 & 0.689 & & 2007\_014555 & 4426.6056 & 0.696 & & 2008\_011780 & 4666.7931 & 0.522\\ 
2008\_011781 & 4666.8002 & 0.527 & & 2008\_011782 & 4666.8072 & 0.531 & & 2008\_011783 & 4666.8153 & 0.536\\ 
2008\_011784 & 4666.8223 & 0.541 & & 2008\_011785 & 4666.8294 & 0.545 & & 2008\_011786 & 4666.8370 & 0.550\\ 
2008\_011787 & 4666.8440 & 0.554 & & 2008\_011788 & 4666.8511 & 0.559 & & 2008\_011902 & 4667.7799 & 0.146\\ 
2008\_011903 & 4667.7870 & 0.151 & & 2008\_011904 & 4667.7940 & 0.155 & & 2008\_011905 & 4667.8011 & 0.159\\ 
2008\_011906 & 4667.8081 & 0.164 & & 2008\_011907 & 4667.8151 & 0.168 & & 2008\_011908 & 4667.8222 & 0.173\\ 
2008\_011909 & 4667.8292 & 0.177 & & 2008\_011910 & 4667.8362 & 0.182 & & 2008\_012027 & 4668.7396 & 0.753\\ 
2008\_012028 & 4668.7466 & 0.757 & & 2008\_012029 & 4668.7537 & 0.762 & & 2008\_012030 & 4668.7607 & 0.766\\ 
2008\_012031 & 4668.7677 & 0.770 & & 2008\_012032 & 4668.7748 & 0.775 & & 2008\_012033 & 4668.7818 & 0.779\\ 
2008\_012034 & 4668.7888 & 0.784 & & 2008\_012035 & 4668.7958 & 0.788 & & 2008\_012036 & 4668.8029 & 0.793\\ 
2008\_012037 & 4668.8099 & 0.797 & & 2008\_012038 & 4668.8169 & 0.802 & & 2008\_012040 & 4668.8249 & 0.807\\ 
2008\_012041 & 4668.8319 & 0.811 & & 2008\_012042 & 4668.8390 & 0.816 & & 2008\_012043 & 4668.8460 & 0.820\\ 
2008\_012044 & 4668.8530 & 0.824 & & 2008\_012045 & 4668.8601 & 0.829 & & 2009\_003452 & 4880.9971 & 0.924\\ 
2009\_003453 & 4881.0041 & 0.929 & & 2009\_003454 & 4881.0112 & 0.933 & & 2009\_003455 & 4881.0182 & 0.938\\ 
2009\_003456 & 4881.0252 & 0.942 & & 2009\_003457 & 4881.0323 & 0.947 & & 2009\_003459 & 4881.0408 & 0.952\\ 
2009\_003460 & 4881.0478 & 0.956 & & 2009\_003461 & 4881.0549 & 0.961 & & 2009\_003462 & 4881.0619 & 0.965\\ 
2009\_003463 & 4881.0689 & 0.970 & & 2009\_003464 & 4881.0760 & 0.974 & & 2009\_003465 & 4881.0832 & 0.979\\ 
2009\_003466 & 4881.0903 & 0.983 & & 2009\_003467 & 4881.0973 & 0.988 & & 2009\_007464 & 4952.8377 & 0.336\\ 
2009\_007465 & 4952.8448 & 0.341 & & 2009\_007466 & 4952.8518 & 0.345 & & 2009\_007467 & 4952.8588 & 0.350\\ 
2009\_007468 & 4952.8659 & 0.354 & & 2009\_007469 & 4952.8729 & 0.358 & & 2009\_007470 & 4952.8802 & 0.363\\ 
2009\_007471 & 4952.8873 & 0.367 & & 2009\_007472 & 4952.8943 & 0.372 & & 2009\_007473 & 4952.9013 & 0.376\\ 
2009\_007474 & 4952.9084 & 0.381 & & 2009\_007475 & 4952.9154 & 0.385 & & 2009\_007477 & 4952.9249 & 0.391\\ 
2009\_007478 & 4952.9320 & 0.396 & & 2009\_007479 & 4952.9390 & 0.400 & & 2009\_007480 & 4952.9460 & 0.405\\ 
2009\_007481 & 4952.9531 & 0.409 & & 2009\_007482 & 4952.9601 & 0.414 & & 2009\_007483 & 4952.9671 & 0.418\\ 
2009\_007484 & 4952.9742 & 0.422 & & 2009\_007485 & 4952.9812 & 0.427 & & 2009\_007486 & 4952.9882 & 0.431\\ 
2009\_007487 & 4952.9953 & 0.436 & & \\
\sidehead{CFHT data}
852275 & 3901.0630 & 0.491 & & 852692 & 3905.0518 & 0.012 & &
\enddata

\end{deluxetable}

\clearpage

\begin{deluxetable}{rrrr}
\tablecaption{Magnetic field observations of \hdnum.\label{magnetic_data}}
\tablewidth{0pt}
\tablehead{
\colhead{HJD} & $B_l$ & $\sigma$ & \colhead{$\phi$} \\
\colhead{(-245\,0000)} & (G) & (G) &
}
\startdata
4487.1038 &    662  &  182 & 0.938 \\
4632.9152 & -1518  &  168 & 0.108 \\
4633.8834 &  1857   & 109 & 0.720 \\
4634.8814 & -2181  & 394 & 0.351 \\ 
4636.8170 &  1411  &  277 & 0.574 \\
4639.9378 &   -564  &  410 & 0.547 \\
4640.7843 & -1317  &  206 & 0.082 \\
4657.8022 &  1921  &  271 & 0.839 \\
4657.9369 &   1462  &  255 & 0.925 \\
4658.7856 &   -647  &  169 & 0.461 \\
4658.9200 &    203  &  226 & 0.546 \\ 
4659.9364 &  -1729  &  173 & 0.189 \\
4661.8954 &   -706  &  192 & 0.427 \\ 
4716.7864 & -1291  & 220 & 0.124 \\
4717.6840  &  1988  & 184 & 0.692 \\
4936.9993  & -1713  & 183 & 0.325
\enddata
\end{deluxetable}

\clearpage

\begin{deluxetable}{cc}
\tabletypesize{\scriptsize}
\tablecaption{Observed and Derived Properties of \hdnum.\label{properties}}
\tablewidth{0pt}
\tablehead{
\colhead{Parameter} & \colhead{Value}
}
\startdata
Sp. Type & B5\,IV\tablenotemark{a}\\
$V$ & 6.404\tablenotemark{a} \\
$B-V$ & -0.155\tablenotemark{a} \\
$U-B$ & -0.729\tablenotemark{a} \\
$\pi$ & $3.42\pm0.30$\,mas\tablenotemark{b} \\
d & $292^{+28}_{-24}$\,pc\tablenotemark{b} \\
$E(B-V)$ & 0.051 \\
$M_V$ & $-1.08^{+0.18}_{-0.20}$ \\
$T_{eff}$ & $16\,000 \pm 1000$ K \\
$\log{g}$ & $4.0\pm0.1$ \\
$BC$ & $-1.38\pm0.16$ \\
$\log{L/L_{\odot}}$ & $2.88 \pm 0.14$ \\
$R/R_{\odot}$ & $3.61^{+0.56}_{-0.45}$ \\
$v\sin{i}$ & $105\pm10$\,km\,s$^{-1}$ \\
RV & $-16\pm2$\,km\,s$^{-1}$\\
$i$ & $\ge 52$\degr; adopted 85\degr \\
$\beta$ & $\approx 77$\degr \\
$B_{\rm d}$ & $>7.0\,$kG\\
$B_{\rm eq}$ & $>2.7\,$kG\\
P & $1.^{\!\!\rm{d}}5819840\pm0.^{\!\!\rm{d}}0000030$\\
E & $2454496.6938 \pm 0.0017$ \\
\enddata
\tablenotetext{a}{\citet{bsc}}
\tablenotetext{b}{\cite{vanleeuwen07}}
\end{deluxetable}


\begin{thebibliography}{}
\bibitem[Abt(2002)]{abt02} Abt, H.A., Levato, H., \& Grosso, M. 2002, \apj, 573, 359
\bibitem[Allen(1976)]{allen76} Allen, C.W. 1976, Astrophysical Quantities (3rd ed.; London: Athlone Press)
\bibitem[Babel \& Montmerle(1997a)]{babel97a} Babel, J., \& Montmerle, T., 1997a, \aap, 323, 121
\bibitem[Babel \& Montmerle(1997b)]{babel97b} Babel, J., \& Montmerle, T., 1997b, \aap, 485, L29
\bibitem[Balona(1994)]{balona94}Balona, L.A. 1994, \mnras, 268, 119
\bibitem[Catanzaro et al.(1999)]{catanzaro99} Catanzaro, G., Leone, F. \& Catalano, F.A. 1999, \aaps, 134, 211
\bibitem[Donati et al.(1997)]{donati97} Donati J.-F., et al. 1997, \mnras, 291, 658
\bibitem[Drake, Wade \& Linsky(2006)]{drake06} Drake, S.A., Wade, G.A. \& Linsky, J.L. 2006, in proceedings of The X-ray Universe, ed. A. Wilson (ESA SP-604), 73
\bibitem[ESA(1997)]{esa97} ESA SP-1200, 1997
\bibitem[Groote \& Hunger(1976)]{groote76} Groote, D., \& Hunger, K. 1976, \aap, 52, 303
\bibitem[Groote \& Hunger(1997)]{groote97} Groote, D., \& Hunger, K. 1997, \aap, 319, 250
\bibitem[Hoffleit \& Warren(1991)]{bsc} Hoffleit D., \& Warren Jr, W.H. 1991, The Bright Star Catalogue, 5th Revised Edition
\bibitem[Hubeny \& Lanz(1995)]{hubeny95} Hubeny, I., \& Lanz, T. 1995, \apj, 439, 875
\bibitem[Huang \& Gies(2008)]{huang08} Huang, W., \& Gies, D.R. 2008, \apj, 683, 1045
\bibitem[Krti\u{c}ka et al.(2007)]{krticka07} Krti\u{c}ka, j., Mikul\'{a}\u{s}ek, Z., Zverko, J., \& \u{Z}i\u{z}\u{n}ovsk\'{y}, J. 2007, \aap, 470, 1089
\bibitem[Landstreet(1992)]{landstreet92} Landstreet, J.D. 1992, \aapr, 4, 38
\bibitem[Lanz \& Hubeny(2007)]{lanz07} Lanz, T., \& Hubeny, I. 2007, \apjs, 169, 83
\bibitem[Leone et al.(2010)]{leone10} Leone, F., Bohlender, D.A., Bolton, C.T., Buemi, C., Catanzaro, G., Hill, G.M., \& Stift, M.J. 2010, \mnras, 401, 2739
\bibitem[Mikul\'{a}\u{s}ek et al.(2008)]{mikulasek08} Mikul\'{a}\u{s}ek, Z., et al. 2008, \aap, 485, 585
\bibitem[Molenda-Zakowicz \& Polubek(2004)]{molenda04} Molenda-Zakowicz, J., \& Polubek, G. 2004, \actaa, 54, 281
\bibitem[Moon \& Dworetsky(1985)]{moon85} Moon, T.T., \& Dworetsky, M.M. 1985, \mnras, 217, 305
\bibitem[Nissen(1974)]{nissen74} Nissen, P.E. 1974, \aap, 36, 57
\bibitem[Oksala et al.(2011)]{oksala11} Oksala, M.,E., Wade, G.A., Townsend, R.H.D., Kochukhov, O., \& Owocki, S. 2011, in proceedings of IAU Symposium 272, Active OB Stars, in press
\bibitem[Pedersen(1976)]{pedersen76} Pedersen, H. 1976, \aap, 49, 217
\bibitem[Preston(1967)]{preston67} Preston, G.W. 1967, \apj, 150, 547
\bibitem[Shore(1987)]{shore87} Shore, S.N. 1987, \aj, 94, 731
\bibitem[Shore \& Brown(1990)]{shore90} Shore, S.N., \& Brown, D.N. 1990, \apj, 365, 665
\bibitem[Short \& Bolton(1994)]{short94} Short, C.I., \& Bolton, C.T. 1994, in proceedings of IAU Symposium 162, Pulsation, Rotation, and Mass Loss in Early-type Stars, ed. L.A. Balona, H.F. Henrichs, \& J.L. Le Contel (Dordrecht, Kluwer), 171  
\bibitem[Smith et al.(2006)]{smith06} Smith, M.A., Wade, G.A., Bohlender, D.A., \& Bolton, C.T. 2006, \aap, 458, 581
\bibitem[Smith \& Bohlender(2007)]{smith07} Smith, M.A., \& Bohlender, D.A. 2007, \aap, 75, 1027
\bibitem[Townsend et al.(2010)]{townsend10} Townsend, R.H.D., Oksala, M.E., Cohen, D.H., Owocki, S.P., ud-Doula, A. 2010, \apj, 714, 318L
\bibitem[Townsend \& Owocki(2005)]{townsend05a} Townsend, R.H.D., \& Owocki, S.P. 2005, \mnras, 357, 251
\bibitem[Townsend, Owocki, \& Groote(2005)]{townsend05b} Townsend, R.H.D., Owocki, S.P., \& Groote, D. 2005, \apj, 630, L81
\bibitem[Townsend, Owocki, \& ud-Doula(2007)]{townsend07} Townsend, R.H.D., Owocki, S.P., \& ud-Doula, A. 2007, \mnras, 382, 139
\bibitem[ud-Doula \& Owocki(2002)]{uddoula02} ud-Doula, A., \& Owocki, S.P. 2002, \apj, 576, 413
\bibitem[van Leeuwen(2007)]{vanleeuwen07} van Leeuwen, F. 2007, \aap, 474, 653
\bibitem[Wade et al.(2006)]{wade06} Wade, G.A, et al. 2006, \aap, 458, 569
\end{thebibliography}
\end{document}